# Using Total Margin of Error to Account for Non-Sampling Error in Election Polls: The Case of Nonresponse

Jeff Dominitz
Department of Economics, Justice, and Society – NORC at the University of Chicago

and

Charles F. Manski
Department of Economics and Institute for Policy Research, Northwestern University

October 2024

Abstract

The potential impact of non-sampling errors on election polls is well known, but measurement has focused on the *margin of sampling error*. Survey statisticians have long recommended measurement of total survey error by mean square error (MSE), which jointly measures sampling and non-sampling errors. We think it reasonable to use the square root of maximum MSE to measure the *total margin of error* (TME). Measurement of TME should encompass both sampling error and all forms of non-sampling error. We suggest that measurement of TME should be a standard feature in the reporting of polls. To provide a clear illustration, and because we believe the exceedingly low response rates commonly obtained by election polls to be a particularly worrisome source of potential error, we demonstrate how to measure the potential impact of nonresponse using the concept of TME. We first show how to measure TME when a pollster lacks any knowledge of the candidate preferences of nonrespondents. We then extend the analysis to settings where the pollster has partial knowledge that bounds the preferences of non-respondents. In each setting, we derive a simple poll estimate that approximately minimizes TME—a *midpoint estimate*—and compare it to a conventional poll estimate.

The views expressed in this paper are those of the authors and do not necessarily represent those of NORC at the University of Chicago or Northwestern University. We have benefitted from the opportunity to present this work at the University of Chicago. We are grateful for the comments of Lisa Blumerman, Josh Clinton, Jamie Druckman, and Andrew Gelman.

1. Introduction

Since the middle-1900s, political polling has sought to adhere to an ideal promoted in the statistical literature on survey research: A population of interest is specified, a random sample of persons are drawn from the population, the sampled persons respond to the questions posed, and they do so truthfully and accurately. However, the ideal is essentially never achieved in practice, creating difficulties in interpreting polling data. Polling professionals have long been aware of these difficulties and have often called attention to them. See, for example, the review article by Prosser and Mellon (2018) who find (p. 757): "the level of error has always been substantially beyond that implied by stated margins of error."

Two serious problems persist, one internal to polling research and the other external. The internal problem is that, while polling professionals often mention the issues that impede interpretation of polling data, they do not quantify the implications when reporting poll results, with the exception of sampling imprecision represented by what has been termed the "margin of error" or "margin of sampling error".[1] The external problem is that media reports regularly downplay or ignore the interpretative issues. The internal problem may contribute to the external one. The fact that polling professionals do not measure and report the non-sampling uncertainties in polling encourages journalists to report findings that ignore these uncertainties.

Survey statisticians and researchers have long documented many possible sources of *total survey error* that may arise in polls, including sampling, nonresponse, measurement, and coverage error. See Groves and Lyberg (2010) for a historical review. Among the many problems that make polling less than ideal in practice, perhaps the most egregious is that only a small subset of the sampled persons respond to the questions posed and there is no basis to think that nonresponse is random. For ease of exposition, we focus attention on this source of non-sampling error in election polls.

---

[1] The precise meaning of this term is not always clear. Shirani-Mehr et al. (2018) characterize its common application in election polls as follows (p. 607): "95% margins of error based on the formula for simple random sampling (SRS)—for example, ±3.5 percentage points for an election survey with 800 people." U.S. Census Bureau (2020) defines the margin of error for the American Community Survey (ACS) as follows (p. 80): "The margin of error is the measure of sampling error published with each ACS estimate. A margin of error is the difference between an estimate and its upper or lower confidence bounds. Confidence bounds can be created by adding the margin of error to the estimate (for an upper bound) and subtracting the margin of error from the estimate (for a lower bound). All published margins of error for the ACS are based on a 90 percent confidence level."

Documenting the historical trend towards decreasing response rates, Prosser and Mellon (2018) write (p. 758): "The cornerstone of survey research is probability sampling . . . The primary challenge . . .is the decline in the willingness of people to participate in surveys. Although response rates to most commercial surveys are secret, we know that response rates to face-to-face and telephone surveys have declined precipitously from academic and government surveys."[2] These authors then consider the relationship between response rate and response bias, writing (p. 762): "Non-response bias is not a function of the response rate, but of the interaction between relevant characteristics and the likelihood of responding to a survey." Later considering response bias in more depth, they remark (p. 771): "By far the most common cause of polling error is unrepresentative samples."

Prosser and Mellon observed that, hoping to eliminate response bias, polling analysts commonly weight the available data. They cautioned that weighting can solve only one of two response problems, writing (p. 772):

> "Non-response bias can take two forms. (1) People in different demographic groups may be less likely to answer a survey. (2) Some types of people within demographic groups may be less likely to take a survey. Provided accurate population-level demographic targets are available, the first type of non-response bias is easily dealt with by weighting demographic groups to their correct proportions. The second type of non-response bias is more problematic, and can only be corrected if the correlates of non-response are known, and population level targets are available. . . . Survey researchers have long known that the ability of weighting to correct for survey bias rests on the assumption that respondents mirror non-respondents within weighting categories."

The AAPOR Task Force on Pre-Election Polling (Clinton et al., 2021) called into question the credibility of assuming that respondents mirror non-respondents within categories. See Clinton et al. (2022) for further discussion. Analysis in Horowitz and Manski (1998) showed that conventional weighting of survey responses may exacerbate response bias when nothing is known about the nature of nonresponse.

---

[2] For example, Keeter et al. (2017) find that "the response rates for telephone polls like those conducted for Pew Research Center" fell from 36 percent in 1997 to 9 percent in 2016.

The precipitous decline in response rates to polling surveys is illustrated well by the prominent New York Times/Siena College Poll, which makes public its response rate. An article on survey methodology on the *New York Times* website states[3]: "Often, it takes many attempts to reach some individuals. In the end, fewer than 2 percent of the people our callers try to reach will respond." Specific numbers were cited in a *New York Times* article on May 13, 2024, reporting on a recent poll in six 'battleground" states. Nate Cohn, the newspaper's chief political analyst, wrote (Cohn, 2024): "We spoke with 4,097 registered voters in Arizona, Georgia, Michigan, Nevada, Pennsylvania and Wisconsin from April 28 to May 9, 2024. . . . For this set of polls, we placed nearly 500,000 calls to about 410,000 voters." Thus, the response rate in this poll was approximately 0.01.

One might anticipate that, with responses from only one percent of the registered voters whom the pollsters attempted to contact in spring 2024, the *New York Times* would be very cautious in interpreting the polling data. Not so. The headline of the May 13 article was "Trump Leads in Five Key States." The first sentence of Cohn's text was "Donald J. Trump leads President Biden in five crucial battleground states, a new set of polls shows." The response rate was only mentioned briefly near the end of the article, with no commentary on response bias.

The review article by Prosser and Mellon (2018) exemplifies the internal problem mentioned above. Polling professionals have verbally recognized the potential for response bias to impede interpretation of polling data, but they have not quantified the implications. The *New York Times* reporting in Cohn (2024) exemplifies the external problem. Media coverage of polls downplays or ignores response bias. The internal problem likely contributes to the external one. When they compute the margin of error for a poll, polling professionals only consider sampling imprecision, not the non-sampling error generated by response bias.

---

[3] https://www.nytimes.com/article/times-siena-poll-methodology.html#link-6a942958, accessed June 22, 2024.

Media outlets parrot this margin of sampling error, whose magnitude is usually small enough to give the mistaken impression that polls provide reasonably accurate estimates of public sentiment.

Survey statisticians have long recommended measurement of the total survey error of a sample estimate by its mean square error (MSE), where MSE is the sum of variance and squared bias.[4] MSE jointly measures sampling and non-sampling errors. Variance measures the statistical imprecision of an estimate. Bias stems from non-sampling errors, including non-random nonresponse. Extending the conventional language of polling, we think it reasonable to use the square root of maximum MSE to measure the *total margin of error* (TME).

In practice, statisticians have focused on variance alone. Groves and Lyberg (2010) state (p. 868): "The total survey error format forces attention to both variance and bias terms. . . . . . Most statistical attention to surveys is on the variance terms—largely, we suspect, because that is where statistical estimation tools are best found." In a study that attempted to estimate MSE based on observed differences between poll results and election outcomes, Shirani-Mehr et al. (2018) stated (p. 608): "In contrast to errors due to sample variance, it is difficult - and perhaps impossible - to build a useful and general statistical theory for the remaining components of total survey error."[5] An early exception to the focus on variance alone was suggested by Cochran, Mosteller, and Tukey (1954). However, their exploratory work was not followed up subsequently.

The fact that polling professionals do not quantify potential non-sampling error, including response bias, in the reporting of poll results would be understandable in the absence of appropriate methodology to perform the measurement. However, such methodology does exist, drawing on and extending our previous research in Dominitz and Manski (2017). This paper explains, focusing on nonresponse error in two-candidate elections. It is straightforward in principle, but more complex mathematically, to extend the analysis to account for other sources of non-sampling error and to consider multi-candidate elections. We do not perform these extensions here in the interest of simplicity of exposition.

---

[4] For an early example, see Hansen, Hurwitz, and Bershad (1959).

[5] When Shirani-Mehr et al. (2018) refer to statistical theory, they presumably have in mind a statistical theory that seeks to quantify imprecision ex ante across repeated samples. They motivate their study of observed differences using a sampling model that generates repeated polls, applying a random-coefficients model often used in meta-analysis. This a coherent form of statistical analysis, but the sampling process assumed in this type of meta-analysis is based on a concept of repeated random sampling of empirical studies – in this case, polls – drawn from a hypothetical population of potential studies. This idea has long been controversial. See Manski (2020) for discussion.

Section 2 summarizes the relevant previous research, showing how to jointly measure statistical imprecision and potential response bias when a pollster lacks any knowledge of the candidate preferences of non-respondents. Section 3 extends the analysis to settings where the pollster has partial knowledge that bounds the preferences of non-respondents. We also consider use of background information on respondents, which pollsters commonly employ to compute weighted estimates of candidate preferences.

In each case, we derive the polling point estimate that minimizes TME—a *midpoint estimate* of the form originally derived in Dominitz and Manski (2017)—and compare it to a conventional poll estimate. In the baseline case where the pollster has no knowledge of nonresponse, we find that the optimal midpoint estimate shrinks the conventional estimate of the spread in reported vote shares between candidates towards zero. In other cases, differences between the estimates depend on the nature of knowledge of nonresponse.

We utilize results from a recent presidential election poll to illustrate the applicability of our findings and to demonstrate how different forms of partial knowledge translate into differences in the TME. We identify conditions under which the conventional polling estimate coincides with the midpoint estimate that is optimal with respect to maximum MSE, noting that the TME exceeds the margin of sampling error whenever the pollster has less than complete knowledge of the nature of nonresponse.

Section 4 concludes with a call not only to introduce TME as a standard feature in the reporting of poll results but also to expand the application of this methodology to additional sources of non-sampling error. We illustrate how to account for potential coverage error arising from, for example, differences between the population of "likely voters" and the population of those who ultimately vote.

## 2. Polling With No Knowledge of Nonresponse

### 2.1. Partial Identification and Maximum Response Bias

Measurement of potential response bias has been central to the econometric literature on partial identification that has developed from the late 1980s onward. Reviews of this body of work include Manski (2003, 2007), Tamer (2010), and Molinari (2020).

To explain the applicability of this literature to polling, we consider in this section a simple setting where the objective is to learn the probability $P(y = 1)$ that a binary outcome y takes the value 1 in a population of interest. In the context of polling, we have in mind a two-candidate race. Then y is the preference that a member of the population currently holds for one candidate rather than the other.

Let $z = 1$ if a member of the population would agree to participate in the poll and $z = 0$ if not. The response rate $P(z = 1)$ is the fraction of the population who, if sampled, would agree to participate in the poll (interview response) and respond to the specific polling question (item response). Manski (1989) observed that, by the Law of Total Probability,

(1) $\quad P(y = 1) = P(y = 1|z = 1)P(z = 1) + P(y = 1|z = 0)P(z = 0).$

Random sampling of the population point-identifies $P(z)$ and $P(y = 1|z = 1)$, the candidate preference of respondents to the poll. The sampling process is uninformative about $P(y = 1|z = 0)$, the candidate preference of non-respondents. If nothing is known about the candidate preferences of non-respondents, the latter probability can take any value in the unit interval. Hence, polling with an unlimited sample size reveals that $P(y = 1)$ lies in the interval

(2) $\quad [P(y = 1|z = 1)P(z = 1), \ P(y = 1|z = 1)P(z = 1) + P(z = 0)].$

This interval is called the *identification region* for P(y = 1), or the *identified set*. Note that the width of the interval is P(z = 0), the fraction of the population who would not respond to the poll if asked.

The interval (2) shows what polling with unlimited sample size reveals about P(y = 1). In this idealized setting, there is no statistical imprecision in estimation of P(y = 1|z = 1) and P(z = 1). Nonresponse bias is the only problem. Dominitz and Manski (2017) showed that the estimate of P(y = 1) that minimizes maximum squared bias is the midpoint of interval (2); that is, P(y = 1|z = 1)P(z = 1) + ½P(z = 0). The resulting value of maximum squared bias is ¼P(z = 0)². Thus, when the interval midpoint is the reported result of a poll with no knowledge of nonresponse and unlimited sample size, TME is simply ½P(z = 0).

It is of interest to compare the interval midpoint with the alternative estimate P(y = 1|z = 1). With unlimited sample size, the latter estimate is the conventional one reported by pollsters, the candidate preference of respondents to the poll. The former estimate, the interval midpoint, lies between the latter one and the value ½. Thus, the former estimate shrinks the spread in reported vote shares between the candidates towards zero. The latter estimate has zero bias if nonresponse is random; that is, if P(y|z = 0) = P(y|z = 1). It has non-zero bias otherwise. Bias is maximized when the candidate preference of non-respondents is as different as possible from the candidate preference of respondents; that is, P(y = 1|z = 0) = 0 if P(y = 1|z = 1) ≥ ½, or P(y = 1|z = 0) = 1 if P(y = 1|z = 1) < ½. The maximum squared bias is therefore the maximum of [P(y = 1|z = 1)P(z = 0)]² and [P(y = 0|z = 1)P(z = 0)]². This maximum squared bias equals ¼P(z = 0)² if P(y = 1|z = 1) = ½ and is larger otherwise.

Thus, maximum squared bias of the conventional estimate coincides with that of the interval midpoint in the special case where respondent preferences are equally split between the two candidates. In that case, both the conventional estimate and the interval midpoint take the value ½. Otherwise, the conventional estimate is further from ½ and has larger maximum squared bias than the interval midpoint.

2.2. Use of MSE to Measure Uncertainty Arising from Partial Identification and Statistical Imprecision

Dominitz and Manski (2017) studied a more realistic setting with finite sample size. We supposed that the population response rate $P(z = 1)$ is known.[6] We focused on the fact that one must estimate $P(y = 1|z = 1)$ using data from the finite random sample of persons who agree to participate in the poll and who answer the polling question of interest. Let m denote the fraction of the respondents who report $y = 1$. Then a natural sample estimate of $P(y = 1|z = 1)$ is m. This yields $[m \cdot P(z = 1), m \cdot P(z = 1) + P(z = 0)]$ as a natural estimate of interval (2). A natural computable point estimate of $P(y = 1)$ is the midpoint of this interval; that is, $m \cdot P(z = 1) + \frac{1}{2} P(z = 0)$.

Dominitz and Manski (2017) derived the maximum MSE of this midpoint estimate, conditioning on the size N of the sample of respondents.[7] The result turns out to be simple, being $\frac{1}{4}[P(z = 1)^2/N + P(z = 0)^2]$. The term $\frac{1}{4} P(z = 1)^2/N$ is maximum variance. The term $\frac{1}{4} P(z = 0)^2$ is maximum squared bias, just as it is in the case above for the interval midpoint with unlimited sample size. The derivation shows that maximum MSE occurs when these components are maximized separately, with variance maximized when $P(y = 1|z = 1) = \frac{1}{2}$ and squared bias maximized when the distribution of candidate preferences among non-respondents is degenerate, such that $P(y = 1|z = 0) = 0$ or $P(y = 1|z = 0) = 1$.

Observe that maximum variance goes to zero with increasing sample size at rate 1/N, but maximum bias does not vary with sample size. The latter property is inevitable in the absence of knowledge of the candidate preferences of non-respondents. The available data are uninformative about $P(y = 1|z = 0)$, whatever the sample size may be.

The lesson is clear. In the absence of knowledge of the candidate preferences of non-respondents, a pollster wanting to give a point estimate of $P(y = 1)$ would do better to report $m \cdot P(z = 1) + \frac{1}{2} P(z = 0)$ rather than m. The pollster should not report the margin of sampling error, which ignores response bias. Instead, it would be appropriate to report the TME as the square root of maximum mean square error; that is, $\frac{1}{2}[P(z = 1)^2/N + P(z = 0)^2]^{\frac{1}{2}}$.

---

[6] Observation of data from one poll of finite size only enables estimation of the response rate. However, if the response rate is stable across polls, data may be combined across polls to approximate it well.

[7] Conditioning on N is appropriate if the data collection protocol calls on the pollster to sample members of the population until one obtains a predetermined number N of respondents. Then N is fixed by design. Some protocols differ from this, instead fixing the total number of persons for whom contact is attempted. Then the number N of respondents is random. In this setting, the analysis of maximum MSE reported by Dominitz and Manski remains valid as an instance of inference conditional on N, which is an ancillary statistic. Computation of maximum MSE unconditional on N is possible but more complex.

2.3. Illustration

Consider the results of the New York Times/Siena College (NYT/SC) presidential election poll conducted among 1,532 registered voters nationwide from June 28 to July 2, 2024.[8] Regarding nonresponse, the reported results include this statement: "For this poll, we placed more than 190,000 calls to more than 113,000 voters." Thus, $P(z = 1) \cong 0.0136$.

We focus here on the following poll results:

| (Includes leaners) **If the 2024 presidential election were held today, who would you vote for if the candidates were:** | REGISTERED VOTERS |
|---|---|
| Joe Biden, the Democrat | 41% |
| Donald Trump, the Republican | 49% |
| [VOL] Don't know/Refused | 10% |
| **MARGIN** Trump +9 Calculated using candidate | |
| Number of respondents | 1,532 |
| Percentage of total electorate | 100% |

Regarding sampling imprecision, the reported results include this statement: "The poll's margin of sampling error among registered voters is plus or minus 2.8 percentage points."

Shirani-Mehr et al. (2018) characterize standard practices in the reporting of poll results. Regarding vote share, they write (p. 609): "As is standard in the literature, we consider *two-party* poll and vote share: we divide support for the Republican candidate by total support for the Republican and Democratic candidates, excluding undecided and supporters of any third-party candidates."

---

[8] https://www.nytimes.com/interactive/2024/07/03/us/elections/times-siena-poll-registered-voter-crosstabs.html?smid=nytcore-ios-share&referringSource=articleShare&sgrp=c-cb, accessed July 7, 2024.

Let P(y = 1|z = 1) denote the preference for the Republican candidate Donald Trump among respondents, discarding those who volunteer "Don't know" or "Refused." Let m denote the conventional estimate of that preference. Thus, m = 0.49/0.90 = 0.544.

Regarding margin of sampling error, Shirani-Mehr et al. write (p. 608): "Most reported margins of error assume estimates are unbiased, and report 95% confidence intervals of approximately ± 3.5 percentage points for a sample of 800 respondents. This in turn implies the RMSE for such a sample is approximately 1.8 percentage points." This passage suggests that the standard practice for calculating the margin of sampling error assumes random nonresponse and maximum variance, which occurs when P(y = 1|z = 1) = ½. Thus, the formula for a poll's margin of sampling error is $1.96[(.5)(.5)/N]^{1/2}$. With 1,532 respondents to the NYT/SC poll, the margin of sampling error is approximately ± 2.5 percentage points.[9] Thus, the conventional poll result for Donald Trump, the Republican, would be 54.4% ± 2.5%. Assuming that nonresponse is random, the square root of the maximum MSE is about 0.013.

What are the midpoint estimate and the TME for this poll, with no knowledge of nonresponse? Recall that the midpoint estimate is m·P(z = 1) + ½P(z = 0) and the square root of maximum MSE is ½[P(z = 1)²/N + P(z = 0)²]^½. Setting m = 0.544, P(z = 1) = 0.014 and N = 1532, the midpoint estimate is 0.501 and the square root of maximum MSE is 0.493. Thus, the poll result for Trump is 50.1% ± 49.3%.

The finding of such a large TME should not be surprising. With a response rate of just 1.4 percent and no knowledge of nonresponse, little can be learned about P(y = 1) from the poll, regardless of the size of the sample of respondents. Even with unlimited sample size, the TME for a poll with a 1.4 percent response rate remains 49.3%.

---

[9] The reported margin of error for the poll includes a reported design effect of 1.22 "due to survey design and weighting", which increases the margin of error from 2.5 percentage points to $1.96[1.22(.5)(.5)/1532]^{1/2}$ or 2.8 percentage points. Shirani-Mehr et al. (2018) find that polling organizations often ignore such design effects.

3. Polling With Partial Knowledge of Nonresponse

We now demonstrate how assertions regarding partial knowledge of nonresponse reduce the TME. The derivation in Section 2 was agnostic about nonresponse, presuming that nothing is known about the candidate preferences of non-respondents to a poll. Hence, $P(y = 1|z = 0)$ could take any value in the unit interval. A pollster may find it credible to assert some partial knowledge of nonresponse. This section shows how to use such knowledge, extending the analysis of Dominitz and Manski (2017).

We consider two situations. In Section 3.1, the pollster thinks it credible to assume that $P(y = 1|z = 0)$ lies in some predetermined informative interval. In Section 3.2, the pollster thinks it credible to assume that $P(y = 1|z = 0)$ lies within some predetermined distance from $P(y = 1|z = 1)$, the observable candidate preference of respondents.[10] We continue to assume knowledge of the response rate and random sampling of the population. We continue to study maximum MSE conditional on the number N of sampled respondents.

As mentioned in the Introduction, pollsters often hope to eliminate response bias by weighting demographic groups to their correct proportions in the population. In Section 3.3, we consider the possible use of respondent background information to reduce TME.

3.1. Predetermined Bound on $P(y = 1|z = 0)$

A pollster may think it credible to assume that $P(y = 1|z = 0)$ lies in an interval $[\lambda_0, \lambda_1]$, where $0 \leq \lambda_0 \leq \lambda_1 \leq 1$.[11] Then polling with an unlimited sample size reveals that $P(y = 1)$ lies in the interval

---

[10] A pollster may instead think it credible to assume that the candidate preference of interest $P(y = 1)$ lies in some predetermined informative interval. In that case, the intersection of this predetermined interval and the poll interval calculated without using this information is the identification region for $P(y = 1)$. See the discussion in Chapter 3 of Manski (2007).

[11] A Bayesian would go further and specify a specify a subjective distribution over this interval. Here we posit only knowledge of the support of this prior. We can understand the appeal of the Bayesian approach in some contexts. However, polls have classically been analyzed from a frequentist perspective. We adopt that approach. See Little and Gelman (1998) for a Bayesian approach to this problem.

(3)   $[P(y = 1|z = 1)P(z = 1) + \lambda_0 P(z = 0), P(y = 1|z = 1)P(z = 1) + \lambda_1 P(z = 0)]$.

This interval has width $(\lambda_1 - \lambda_0)P(z = 0)$. Thus, the frequency $P(z = 0)$ of nonresponse and the degree $\lambda_1 - \lambda_0$ of uncertainty about the candidate preferences of non-respondents combine multiplicatively to determine the magnitude of the uncertainty about the value of $P(y = 1)$.

As in Section 2, it is easy to see what polling with unlimited sample size reveals about $P(y = 1)$. With unlimited sample size, nonresponse bias is the only problem. The estimate of $P(y = 1)$ that minimizes maximum squared bias is the midpoint of interval (3), namely $P(y = 1|z = 1)P(z = 1) + \frac{1}{2}(\lambda_0 + \lambda_1)P(z = 0)$. The maximum squared bias of this estimate is $\frac{1}{4}(\lambda_1 - \lambda_0)^2 P(z = 0)^2$. When $\lambda_0$ and $\lambda_1$ are symmetric about 0.5, this interval midpoint equals the interval midpoint with no knowledge of nonresponse. For any informative $\lambda_0$ and $\lambda_1$, maximum squared bias is $(\lambda_1 - \lambda_0)^2$ times its value with no knowledge of nonresponse.

---

*Illustration – Extension 1:* Consider again the NYT/SC presidential poll with a response rate of 0.014. Recall that, even with unlimited sample size, the TME is 49.3% with no knowledge of nonresponse. Suppose that it is credible to place a predetermined bound on $P(y = 1|z = 0)$. For instance, suppose one believes that partisan voters constitute roughly two-thirds of the electorate and are pretty evenly split between Democrat and Republican.[12]

---

[12] The basis for this illustration comes from responses to a poll performed by the Pew Research Center, which reported (2024): "About two-thirds of registered voters identify as a partisan, and they are roughly evenly split between those who say they are Republicans (32% of voters) and those who say they are Democrats (33%). Roughly a third instead say they are independents or something else (35%), with most of these voters leaning toward one of the parties." The reported "cumulative response rate" to the poll is 4%.

Then, with non-respondents constituting 98.6 percent of the population, it may be reasonable to assert that P(y = 1|z = 0) lies in the interval [0.3, 0.7]. In this case, with $\lambda_0$ = 0.30 and $\lambda_1$ = 0.70 being symmetric about 0.5, the midpoint of interval (3) equals the midpoint of interval (2) with no knowledge of nonresponse. The maximum of squared bias is ¼(0.7-0.3)²(0.986)² = 0.039 and the TME falls from 49.3% to 19.7%.

---

Now suppose that the response rate P(z = 1) is known, but one must estimate P(y = 1|z = 1) using data from a finite sample of size N who respond to the poll. As in Section 2, a natural estimate of P(y = 1|z = 1) is the fraction m of the sample who report y = 1. Hence, one may estimate P(y = 1) by m·P(z = 1) + ½($\lambda_0$ + $\lambda_1$)P(z = 0).

Extension of the analysis in Dominitz and Manski (2017), Section 3.4 yields the maximum MSE of this estimate, which we now denote as p. Let S be the state space, which indexes all feasible values of [P(y = 1|z = 1), P(y = 1|z = 0)]. In each state s ∈ S, the MSE of p is

(5) $\text{Var}_s(p) + [\text{Bias}_s(p)]^2 = E_s[p - E_s(p)]^2 + [E_s(p) - P_s(y = 1)]^2$,

where

(6) $E_s(p) = P_s(y = 1|z = 1) \cdot P(z = 1) + ½(\lambda_0 + \lambda_1)P(z = 0)$,

(7) $P_s(y = 1) = P_s(y = 1|z = 1)P(z = 1) + P_s(y = 1|z = 0)P(z = 0)$.

Hence,

(8) $\text{Var}_s(p) = E_s\{[m - P_s(y = 1|z = 1)]^2\}P(z = 1)^2 = [P_s(y = 1|z = 1)P_s(y = 0|z = 1)/N]P(z = 1)^2$,

(9)     $[\text{Bias}_s(p)]^2 = [P_s(y = 1|z = 0) - \frac{1}{2}(\lambda_0 + \lambda_1)]^2 P(z = 0)^2.$

The set of feasible values of the unknowns is the Cartesian Product set $[P_s(y = 1|z = 1), P_s(y = 1|z = 0); s \in S] = [0, 1] \times [\lambda_0, \lambda_1]$. $\text{Var}_s(p)$ is maximized when $P_s(y = 1|z = 1) = \frac{1}{2}$, which yields $\text{Var}_s(p) = \frac{1}{4}P(z = 1)^2/N$. $[\text{Bias}_s(p)]^2$ is maximized when $P_s(y = 1|z = 0)$ takes either of its two extreme feasible values, $\lambda_0$ or $\lambda_1$. Thus, $\max_s [\text{Bias}_s(p)]^2 = \frac{1}{4}P(z = 0)^2 \cdot (\lambda_1 - \lambda_0)^2$. Variance and squared bias can be simultaneously maximized, because the state space has the Cartesian Product form. Hence, maximum MSE is

(10)    $\text{Max}_{s \in S} \text{Var}_s(p) + [\text{Bias}_s(p)]^2 = \frac{1}{4}[P(z = 1)^2/N + P(z = 0)^2 \cdot (\lambda_1 - \lambda_0)^2].$

The magnitude of maximum MSE decreases with sample size N and as the width $\lambda_1 - \lambda_0$ of the predetermined bound on $P(y = 1|z = 0)$ shrinks. It decreases with the response rate $P(z = 1)$ if and only if $1/N < (\lambda_1 - \lambda_0)^2$.[13]

---

*Illustration – Extension 2:* Consider again the NYT/SC presidential poll with a response rate of 0.014 and N=1532. Recall that the conventional poll result for Trump would be 54.4% ± 2.5%, while, with no knowledge of nonresponse, the midpoint estimate would be 50.1% with a TME of 49.3%. Now, with knowledge that $P(y = 1|z = 0)$ lies in the interval [0.3, 0.7], the midpoint estimate remains at 50.1%, because the interval is symmetric about 0.5. The maximum MSE is $\frac{1}{4}[(0.014)^2/1532 + (0.986)^2 \cdot (0.4)^2] = 0.039$, and the TME is

---

[13] This relationship arises because $\lambda_1 - \lambda_0$ characterizes the degree of uncertainty about candidate preferences of non-respondents. It is revealing to consider two polar settings. With no knowledge of nonresponse, $\lambda_1 - \lambda_0 = 1$ and maximum MSE decreases with the response rate as shown in Section 2. In contrast, with full knowledge of nonresponse, $\lambda_1 - \lambda_0 = 0$ and maximum bias = 0. In this polar case, increasing the response rate increases maximum variance with no complementary reduction in maximum bias; hence, maximum MSE increases.

19.7%. Note that, as in the original illustration of Section 2.3, the sampling variance has no noticeable impact on the TME, which is also 19.7% in *Extension 1*. Further, note that, with $\lambda_1 - \lambda_0 = 0.4$, the magnitude of maximum MSE decreases with the response rate for any poll with more than 2 respondents.

---

3.2. Predetermined Distance Between $P(y = 1|z = 0)$ and $P(y = 1|z = 1)$

A pollster may think it credible to assume that the unobserved candidate preference $P(y = 1|z = 0)$ of non-respondents does not differ too much from the observed preference $P(y = 1|z = 1)$ of respondents. Thus, the pollster may assume that $\delta_0 \leq P(y = 1|z = 0) - P(y = 1|z = 1) \leq \delta_1$, for specified constants $\delta_0$ and $\delta_1$. For instance, the pollster may believe that the candidate preferences differ symmetrically by no more than $\delta$, such that $-\delta_0 = \delta_1 = \delta$. Or the pollster may believe that candidate preferences among non-respondents systematically differ from the preferences of respondents in a known direction, such that $0 \leq \delta_0 \leq \delta_1$. Regardless, the inequalities $-P(y = 1|z = 1) \leq \delta_0 \leq \delta_1 \leq 1 - P(y = 1|z = 1)$ must hold to render $P(y = 1|z = 0)$ a proper probability. Then polling with an unlimited sample size reveals that $P(y = 1)$ lies in the interval

(11)    $[P(y = 1|z = 1) + \delta_0 P(z = 0),\ P(y = 1|z = 1) + \delta_1 P(z = 0)]$.

This interval has width $(\delta_1 - \delta_0)P(z = 0)$.

As above, with unlimited sample size, nonresponse bias is the only problem. The estimate of $P(y = 1)$ that minimizes maximum squared bias is the midpoint of interval (11), namely $P(y = 1|z = 1) + \frac{1}{2}(\delta_0 + \delta_1)P(z = 0)$. The maximum squared bias of this estimate is $\frac{1}{4}(\delta_1 - \delta_0)^2 P(z = 0)^2$. When $-\delta_0 = \delta_1 = \delta$, this interval midpoint equals the conventional poll estimate $P(y = 1|z = 1)$. In the extreme, with $\delta = 0$, nonresponse is known to be random and maximum squared bias is 0.

*Illustration – Extension 3:* Suppose the pollster believes that "supporters of the trailing candidate may be less likely to respond to surveys" (Shirani-Mehr et al., 2018, p. 608). In the case where Joe Biden, the Democrat, is the "trailing candidate" in the 2024 presidential election campaign, one may interpret this knowledge as follows: $P(y = 1|z = 1) > 0.5$ and $\delta_0 \leq \delta_1 \leq 0$. Suppose $\delta_0 = -0.1$ and $\delta_1 = 0.0$, meaning that the candidate preference of non-respondents is known to differ from the preference of respondents by no more than 10 percentage points. With a NYT/SC poll response rate of 0.014, the maximum squared bias is $\frac{1}{4}(-0.1-0.0)^2(0.986)^2 = 0.0024$ and the TME falls to 4.9%.

Now suppose that the response rate $P(z = 1)$ is known, but one must estimate $P(y = 1|z = 1)$ using data from a sample of size N who respond to the poll. Using m to estimate $P(y = 1|z = 1)$, one may estimate $P(y = 1)$ by $m + \frac{1}{2}(\delta_0 + \delta_1)P(z = 0)$. This estimate equals m if $-\delta_0 = \delta_1$. However, the estimate may lie outside of the [0, 1] interval in some samples if $-\delta_0 \neq \delta_1$.

Consider the maximum MSE of the estimate, which we denote q. Let the state space S index all feasible values of $P(y = 1|z = 1)$ and the difference $P(y = 1|z = 0) - P(y = 1|z = 1)$. In each state s, the MSE of q is

(12) $\quad \text{Var}_s(q) + [\text{Bias}_s(q)]^2 = E_s[q - E_s(q)]^2 + [E_s(q) - P_s(y = 1)]^2$,

where

(13) $\quad E_s(q) = P_s(y = 1|z = 1) + \frac{1}{2}(\delta_0 + \delta_1)P(z = 0)$.

Hence,

(14)    $\text{Var}_s(q) = E_s\{[m - P_s(y = 1|z = 1)]^2\} = [P_s(y = 1|z = 1)P_s(y = 0|z = 1)/N]$,

(15)    $[\text{Bias}_s(q)]^2 = P(z = 0)^2\{\frac{1}{2}(\delta_0 + \delta_1) - [P_s(y = 1|z = 0) - P_s(y = 1|z = 1)]\}^2$.

With $\delta_0$ and $\delta_1$ pre-specified, the feasible values of $P_s(y = 1|z = 1)$ are those that satisfy the inequalities $-P_s(y = 1|z = 1) \leq \delta_0 \leq \delta_1 \leq 1 - P_s(y = 1|z = 1)$ that must hold to render $P_s(y = 1|z = 0)$ a proper probability.

Analysis of maximum MSE is simplest when $P_s(y = 1|z = 1) = \frac{1}{2}$ is feasible.[14] Then $\text{Var}_s(q)$ is maximized at this value, which yields $\text{Var}_s(q) = 1/(4N)$. $[\text{Bias}_s(q)]^2$ is maximized when the candidate preference difference $P_s(y = 1|z = 0) - P_s(y = 1|z = 1)$ takes either of its two extreme feasible values, $\delta_0$ or $\delta_1$. Thus, $\max_{s \in S} [\text{Bias}_s(q)]^2 = \frac{1}{4}P(z = 0)^2 \cdot (\delta_1 - \delta_0)^2$. Variance and squared bias can be simultaneously maximized. Hence, maximum MSE is

(16)    $\text{Max}_{s \in S} \text{Var}_s(q) + [\text{Bias}_s(q)]^2 = \frac{1}{4}[1/N + P(z = 0)^2 \cdot (\delta_1 - \delta_0)^2]$.

The magnitude of the maximum MSE decreases with sample size N, with the response rate $P(z = 1)$, and as $\delta_1 - \delta_0$ shrinks.

---

*Illustration – Extension 4:* Consider again the NYT/SC poll of 1532 registered voters with m = 0.544. Suppose that Biden is the "trailing candidate" whose supporters are less likely to respond to the poll, with $\delta_0 = -0.1$ and $\delta_1 = 0.0$. With a response rate of 0.014, the midpoint estimate is 0.544 + (-.05)(0.986) = 0.495. If $P_s(y = 1|z = 1) = \frac{1}{2}$ is feasible, then the maximum MSE is ¼[1/1532 + $(0.986)^2 \cdot (0.1)^2$] = 0.003. Thus, under the assumption that non-respondents are up to 10 percentage points more likely to express a preference for Biden than are respondents, the midpoint estimate for Trump falls to 49.5% with a TME of 5.1%.[15]

---

[14] Analysis is straightforward when $P_s(y = 1|z = 1) = \frac{1}{2}$ is not feasible. Variance is maximized at the feasible value for $P_s(y = 1|z = 1)$ that is closest to $\frac{1}{2}$. Bias is maximized as in the case where $P_s(y = 1|z = 1) = \frac{1}{2}$ is feasible.

3.3. Estimation of Candidate Preference Using Respondent Background Information

We have thus far considered estimation of candidate preferences using only polling responses. Pollsters commonly observe some background attributes of respondents and sometimes use this information when computing their estimates. In particular, they often weight groups of respondents to their correct proportions in the population, hoping that this will eliminate response bias. As mentioned in the Introduction, the argument for weighting assumes that, within each group, respondents and non-respondents have the same candidate preferences.

To study estimation without this strong assumption, we return to the Law of Total Probability (1) and restate it conditioning on background attributes that are observed for the persons who respond to a poll. Formally, suppose that a pollster observes attributes x for each respondent, where x lies in a finite set X. For example, x may classify the person by race, age, sex, and/or years of schooling. We now restate (1), conditioning on x:

$$(17) \quad P(y = 1) = \sum_{\xi \in X} P(y = 1 | x = \xi) P(x = \xi)$$

$$= \sum_{\xi \in X} [P(y = 1 | x = \xi, z = 1) P(z = 1 | x = \xi) + P(y = 1 | x = \xi, z = 0) P(z = 0 | x = \xi)] \cdot P(x = \xi)$$

---

[15] If, instead, $P_s(y = 1 | z = 1)$ is known to be greater than ½ in this "trailing candidate" scenario, then the TME would still be 5.1% even if the minimum feasible value for $P_s(y = 1 | z = 1)$ were as large as 0.6.

$$= \sum_{\xi \in X} P(y = 1|x = \xi, z = 1)P(z = 1|x = \xi) \cdot P(x = \xi) + \sum_{\xi \in X} P(y = 1|x = \xi, z = 0)P(z = 0|x = \xi) \cdot P(x = \xi)$$

Of the quantities on the right-hand side of (17), random sampling of the population point-identifies $P(y = 1|x = \xi, z = 1)$, $\xi \in X$, the candidate preference of respondents conditional on observed attributes. We will suppose that a pollster uses Census or other available population data to learn the population distribution of x; that is, $P(x = \xi)$, $\xi \in X$. The pollster can then use Bayes Theorem to learn $P(z = 1|x = \xi)$, $\xi \in X$ and, hence, $P(z = 0|x = \xi)$, $\xi \in X$. That is, by Bayes Theorem,

(18) $P(z = 1|x = \xi) = P(x = \xi|z = 1)P(z = 1)/P(x = \xi)$.

Of the quantities on the right-hand side of (18), random sampling of the population point-identifies $P(x = \xi|z = 1)$ and $P(z = 1)$. Hence, $P(z = 1|x = \xi)$ is point-identified given the assumed knowledge of $P(x = \xi)$.

Thus, random sampling of the population, along with prior knowledge of the population distribution of background attributes, point-identifies each quantity in (17) except $P(y = 1|x = \xi, z = 0)$, $\xi \in X$, the candidate preference of non-respondents conditional on observed attributes.

3.3.1. No Knowledge of Nonresponse

Extending our previous analysis, the central problem is that the sampling process is uninformative about $P(y = 1|x = \xi, z = 0)$, $\xi \in X$. If nothing is known about the candidate preferences of non-respondents, each of these probabilities can take any value in the unit interval. Hence, polling with an unlimited sample size only reveals that $P(y = 1)$ lies in the interval (2) as earlier.

Thus, with no knowledge of the candidate preferences of non-respondents, observation of background attributes of poll respondents cannot reduce response bias. Methods that have been advocated to deal with nonresponse, including weighting and imputation, are useless. See Horowitz and Manski (1998) for analysis in depth.

3.3.2. Predetermined Bound on $P(y = 1|x = \xi, z = 0)$

Observation of background attributes of respondents may be useful if a pollster has partial knowledge of the candidate preferences of non-respondents and this knowledge varies with the attributes x. Extending the analysis of Section 3.1, a pollster may think it credible to assume that, for each $\xi \in X$, $P(y = 1|x = \xi, z = 0)$ lies in an interval $[\lambda_{0\xi}, \lambda_{1\xi}]$, where $0 \leq \lambda_{0\xi} \leq \lambda_{1\xi} \leq 1$. Then it follows from (17) that polling with an unlimited sample size reveals that $P(y = 1)$ lies in the interval

$$(19) \quad [\sum_{\xi \in X} P(y = 1|x = \xi, z = 1)P(z = 1|x = \xi) \cdot P(x = \xi) + \sum_{\xi \in X} \lambda_{0\xi} P(z = 0|x = \xi) \cdot P(x = \xi),$$

$$\sum_{\xi \in X} P(y = 1|x = \xi, z = 1)P(z = 1|x = \xi) \cdot P(x = \xi) + \sum_{\xi \in X} \lambda_{1\xi} P(z = 0|x = \xi) \cdot P(x = \xi)].$$

This interval has smaller width than that given in (3) if the bounds $[\lambda_{0\xi}, \lambda_{1\xi}]$, $\xi \in X$ are narrower, on average,[16] than the unconditional bound $[\lambda_0, \lambda_1]$ that the pollster can place on $P(y = 1|z = 0)$ without observation of background attributes.

The estimate of $P(y = 1)$ that minimizes maximum squared bias is the midpoint of interval (19), namely

$$\sum_{\xi \in X} P(y = 1|x = \xi, z = 1)P(z = 1|x = \xi) \cdot P(x = \xi) + \tfrac{1}{2} \sum_{\xi \in X} (\lambda_{0\xi} + \lambda_{1\xi}) P(z = 0|x = \xi) \cdot P(x = \xi).$$

The maximum MSE of this estimate is its maximum squared bias. This occurs when the unknown candidate preferences of non-respondents jointly take on either their lowest or highest possible values; that is, $P(y = 1|x = \xi, z = 0) = \lambda_{0\xi}$, $\xi \in X$ or $P(y = 1|x = \xi, z = 0) = \lambda_{1\xi}$, $\xi \in X$. Maximum MSE is

$$\tfrac{1}{4} \Big[\sum_{\xi \in X} (\lambda_{1\xi} - \lambda_{0\xi}) P(z = 0|x = \xi) \cdot P(x = \xi)\Big]^2.$$

---

[16] "On average" here means when weighted by the relative frequency among non-respondents $P(x = \xi | z = 0)$.

Suppose $P(y = 1|x = \xi, z = 1)$, $\xi \in X$ are not known but are estimated by the fractions of the respondents with attributes $x = \xi$ who report $y = 1$, denoted $m_\xi$. Suppose further that the attribute-specific response rates $P(z = 0|x = \xi)$, $\xi \in X$ are known and that MSE is computed conditional on the sample sizes $N_\xi$, $\xi \in X$, which are ancillary statistics.[17] The estimates $m_\xi$, $\xi \in X$ are statistically independent conditional on $N_\xi$, $\xi \in X$. Hence, extension of the argument in Section 3.1 shows that maximum MSE is the sum of maximum variance

$$\frac{1}{4} \left[ \sum_{\xi \in X} P(z = 1|x = \xi) \cdot P(x = \xi)/N_\xi \right]^2$$

and the above maximum squared bias.

3.3.3. Predetermined Distance Between $P(y = 1|x = \xi, z = 0)$ and $P(y = 1|x = \xi, z = 1)$

Extending the analysis of Section 3.2, a pollster may think it credible to assume that, for each $\xi \in X$, the difference between $P(y = 1|x = \xi, z = 0)$ and $P(y = 1|x = \xi, z = 1)$ lies in a predetermined interval $[\delta_{0\xi}, \delta_{1\xi}]$, where $-P(y = 1|x = \xi, z = 1) \leq \delta_{0\xi} \leq \delta_{1\xi} \leq 1 - P(y = 1|x = \xi, z = 1)$. Then polling with an unlimited sample size reveals that $P(y = 1)$ lies in the interval

(20) $\quad [ \sum_{\xi \in X} P(y = 1|x = \xi, z = 1) P(x = \xi) + \sum_{\xi \in X} \delta_{0\xi} P(z = 0|x = \xi) \cdot P(x = \xi),$

$\quad\quad [ \sum_{\xi \in X} P(y = 1|x = \xi, z = 1) P(x = \xi) + \sum_{\xi \in X} \delta_{1\xi} P(z = 0|x = \xi) \cdot P(x = \xi)]$

This interval has smaller width than that given in (11) if the bounds $[\delta_{0\xi}, \delta_{1\xi}]$, $\xi \in X$ are narrower, on average, than the unconditional bound $[\delta_0, \delta_1]$ that the pollster can place on the difference between $P(y = 1|z = 0)$ and $P(y = 1|z = 1)$ without observation of background attributes.

---

[17] Footnotes 6 and 7 discussed this type of MSE analysis not conditioning on observed respondent background. As noted there, analysis is more complex if response rates must be estimated or if randomness in realized sample sizes is considered.

The estimate of P(y = 1) that minimizes maximum squared bias is the midpoint of interval (20), namely

$$\sum_{\xi \in X} P(y = 1 | x = \xi, z = 1) \cdot P(x = \xi) + \tfrac{1}{2} \sum_{\xi \in X} (\delta_{0\xi} + \delta_{1\xi}) P(z = 0 | x = \xi) \cdot P(x = \xi).$$

The maximum MSE of this estimate is its maximum squared bias. Maximum MSE is

$$\tfrac{1}{4} \left[ \sum_{\xi \in X} (\delta_{1\xi} - \delta_{0\xi}) P(z = 0 | x = \xi) \cdot P(x = \xi) \right]^2.$$

In the special case where $-\delta_{0\xi} = \delta_{1\xi} = \delta_\xi$ for each $\xi \in X$, this interval midpoint equals the conventional poll estimate with population weights $P(x = \xi)$, $\xi \in X$ to account for nonresponse and unlimited sample size. In the extreme case, with $\delta_\xi = 0$ for each $\xi \in X$, maximum squared bias is 0 and nonresponse is known to be random conditional on x, conventionally known as "missing at random." See Section 3.2 of Manski (2007).

Suppose now that $P(y = 1 | x = \xi, z = 1)$, $\xi \in X$, is not known but is estimated by the fraction $m_\xi$ of the $N_\xi$ respondents with attributes $x = \xi$ who report $y = 1$. The midpoint estimate is therefore

$$\sum_{\xi \in X} m_\xi \cdot P(x = \xi) + \tfrac{1}{2} \sum_{\xi \in X} (\delta_{0\xi} + \delta_{1\xi}) P(z = 0 | x = \xi) \cdot P(x = \xi).$$

Given the bounds $[\delta_{0\xi}, \delta_{1\xi}]$, $\xi \in X$, suppose it is feasible that $P(y = 1 | x = \xi, z = 1) = \tfrac{1}{2}$, all $\xi \in X$. Then the maximum variance of this estimate is

$$\tfrac{1}{4} \sum_{\xi \in X} P(x = \xi)^2 / N_\xi.$$

The maximum MSE is the sum of this maximum variance and the above maximum squared bias.

In the special case where $\delta_{0\xi}$ and $\delta_{1\xi}$ are symmetric about zero for all $\xi \in X$, the midpoint estimate reduces to

$$\sum_{\xi \in X} m_\xi \cdot P(x = \xi)$$

This point estimate has the same form as the conventional poll estimate using weights to account for nonresponse, as described by Gelman et al. (2016) as follows (p. 110): "The idea is to partition the population into cells (defined by the cross-classification of various attributes of respondents), use the sample to estimate the mean of a survey variable within each cell, and finally to aggregate the cell-level estimates by weighting each cell by its proportion in the population." That is exactly the procedure here with $\delta_{0\xi}$ and $\delta_{1\xi}$ symmetric about 0 (i.e., $-\delta_{0\xi} = \delta_{1\xi} = \delta_\xi$), cell-level estimates $m_\xi$, and known population proportions $P(x = \xi)$ for each $\xi \in X$. As above, in the extreme case with $\delta_\xi = 0$ for each $\xi \in X$, maximum squared bias is 0 and, therefore, maximum MSE is equal to the maximum squared variance on which margin of sampling error calculations are based.

3.4. Graphical Illustration

To illustrate how the TME relates both to the margin of sampling error and to the extent of knowledge about the nature of nonresponse, we consider a hypothetical poll of 50,000 sampled registered voters that yields a 2% response rate, with 54% percent of respondents expressing a preference for Candidate A and 46% for Candidate B in a two-candidate race. Figure 1 depicts the square root of the maximum MSE—labelled "Total Margin of Error" —for varying assumptions on how much the preferences of non-respondents may differ from respondents, given knowledge that this difference lies in a pre-determined interval of width $2 \times \delta$ that is symmetric about 0. That is, knowledge about the nature of nonresponse may be characterized as follows:

> Assumption – Non-respondent preferences for Candidate A differ from respondent preferences by no more than $\delta \times 100$ percentage points in either direction.

Figure 1. Total Margin of Error for a Hypothetical Poll in a Two-Candidate Race

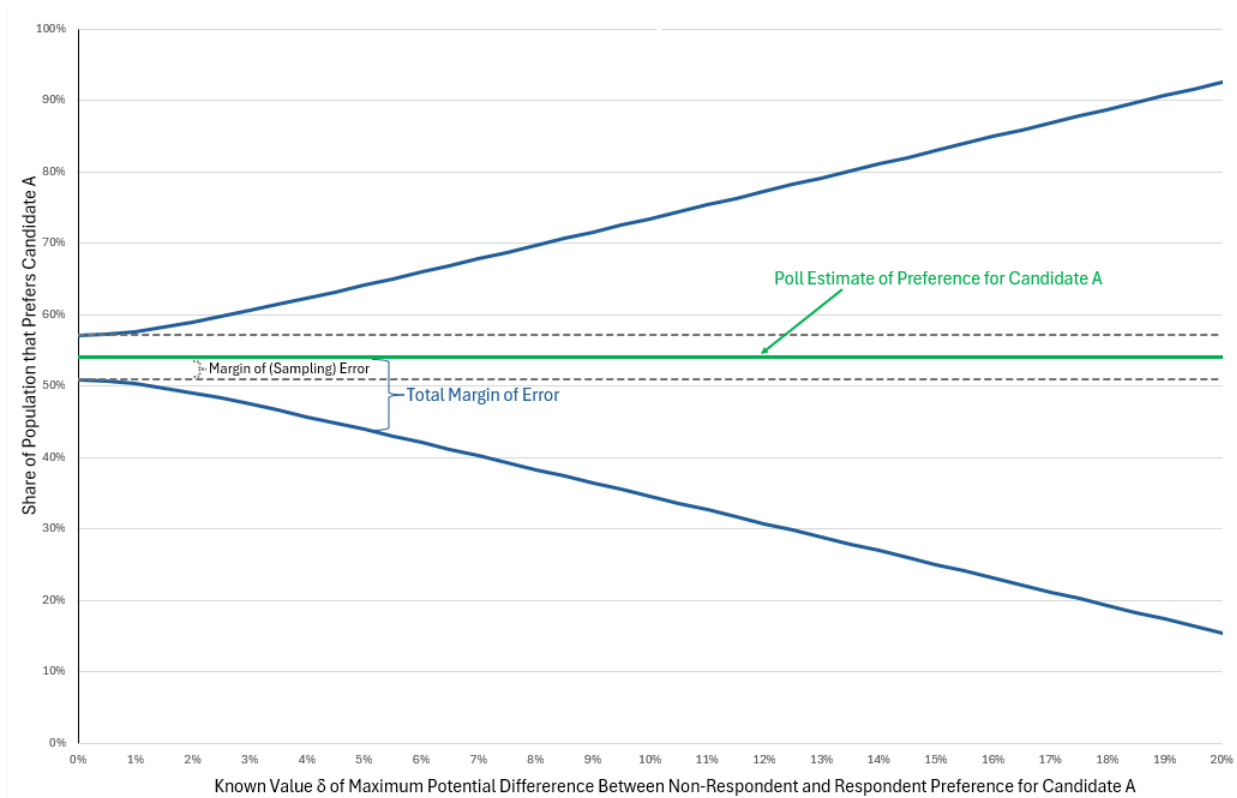

The TME increases as δ increases in Figure 1. The vertical distance between the blue bands depicts the interval defined by the midpoint estimate plus-or-minus the TME for any value of δ. In this setting with -$\delta_0$ = $\delta_1$ = δ, the midpoint estimate coincides with the conventional poll estimate depicted in green. Therefore, the blue bands are centered at the conventional poll estimate in this figure.

When seeking to compare the margin of sampling error to the TME, one complication arises because the former multiplies the square root of maximum variance by a factor such as 1.96 (Shirani-Mehr et al., 2018) or 1.645 (U.S. Census Bureau, 2020),[18] whereas the latter does not. In Figure 1, we calculate the margin of sampling error using the factor 1.96, depicted by the vertical distance between the green line and dashed gray line. Thus, at the far left of the graph where δ = 0 and maximum MSE equals maximum

---

[18] A user manual for the American Community Survey (ACS) reports that "all published margins of error for the ACS are based on a 90 percent confidence level" (U.S. Census Bureau, 2020, p. 80).

25variance, the margin of sampling error is 96 percent larger than the depicted TME. To put the measures on equal footing, one could multiply the square root of maximum MSE by the same factor when defining the TME, in which case the spread of the blue bands would increase proportionally and the two margins would be equal at the y-axis where $\delta = 0$.

4. Conclusion

The potential impact of nonresponse on election polls is well known and frequently acknowledged. Yet the polling profession has not formalized the problem and quantified it ex ante. This paper demonstrates one approach to measuring the potential impact of non-sampling error—in particular, nonresponse error—using the concept of the total margin of error of an election poll.

We briefly called attention to our concerns in a *Roll Call* opinion piece prior to the 2022 midterm elections (Dominitz and Manski, 2022). There we observed that the media response to problems arising from non-sampling error in polls has been to increase the focus on polling averages.[19] We cautioned: "Polling averages need not be more accurate than the individual polls they aggregate. Indeed, they may be less accurate than particular high-quality polls."

We conclude that calculation and reporting of TME should be a standard feature in polling. In principle, the maximum MSE of a poll can be used to quantify not only the potential implications of nonresponse but also additional sources of non-sampling error. One such source is the problem of forecasting turnout. Shirani-Mehr et al. (2018) stressed the importance of potential coverage errors arising from differences between the population of registered voters or of "likely voters", on the one hand, and the population of those who ultimately vote, on the other. To extend the methodology developed here, a pollster would posit credible bounds on such quantities as the proportion of non-voters among the respondents and the proportion of voters among respondents not classified as "likely voters".

---

[19] See Jackson (2015) for a thorough critique of poll averaging.